\documentclass[aps,prl,twocolumn,superscriptaddress,showpacs]{revtex4}

\usepackage{graphicx}
\usepackage{amsmath}

\usepackage{epsfig}
\usepackage{color}
\usepackage{subfigure}

\begin{document}

\title{Quantum-classical correspondence in the wavefunctions of  Andreev billiards}

\author{A.~Korm\'anyos}
\email{kor@complex.elte.hu}
\affiliation{Department of Physics, Lancaster University, Lancaster,
LA1 4YB, UK}
\author{Z.~Kaufmann}
\affiliation{Department of Physics of Complex Systems, E{\"o}tv{\"o}s
University, H-1117 Budapest, P\'azm\'any P{\'e}ter s{\'e}t\'any 1/A, Hungary}
\author{J.~Cserti}
\affiliation{Department of Physics of Complex Systems, E{\"o}tv{\"o}s
University, H-1117 Budapest, P\'azm\'any P{\'e}ter s{\'e}t\'any 1/A, Hungary}
\author{C.~J.~Lambert}
\affiliation{Department of Physics, Lancaster University, Lancaster,
LA1 4YB, UK}

\begin{abstract}
We present a classical and  quantum mechanical study of an 
Andreev billiard with a chaotic normal dot. We demonstrate
  that in general the classical dynamics of 
these normal-superconductor hybrid systems is mixed, thereby indicating
 the limitations of a widely used retracing approximation. 
 We show that the mixed classical dynamics  gives rise to a wealth of wavefunction 
phenomena, including 
periodic orbit scarring and localization of the wavefunction 
onto other classical phase space objects 
such as  intermittent regions and quantized tori. 

\end{abstract}


\maketitle

Wavefunction phenomena in closed and open quantum dot systems have attracted 
much  attention in recent years. Besides the ongoing theoretical interest in a 
simple yet detailed  description of  scarring\cite{ref:heller-scar,ref:vergini-bogomolny}, 
a range of phenomena have been studied including 
the connection between conductance oscillations, transmission resonances of
ballistic semiconductor dots and microwave cavities,   
light emission from dielectric cavities and  the localization of 
the wavefunction onto classical phase space objects\cite{ref:gen-loc}
such as  Kolmogorov-Arnold-Moser (KAM)
islands\cite{ref:moura}, hierarchical regions\cite{ref:backer} 
and periodic orbits\cite{ref:stockmann}.
It has been also  pointed out, that certain states of a closed quantum dot,  
associated with KAM islands or bouncing-ball trajectories, 
actually survive the coupling of the dot to external leads, since  level
broadenings  are  not uniform, and this can give rise to  
measurable transport effects\cite{ref:akis}. 

Systems consisting of 
a ballistic quantum dot coupled to a  superconductor, which are 
commonly called Andreev-billiards (ABs)\cite{ref:andreev-bil,ref:beenakker-rev} 
raise  new questions of quantum-classical 
correspondence addressed by only a few works
\cite{ref:silvestrov,ref:jacquod,ref:goorden,ref:richter,ref:florian} beforehand. 
The key physical process taking place in normal-superconductor (N-S)
hybrid nanostructures  is 
the Andreev reflection\cite{ref:andreev}, whereby   
electron-like quasi-particles
with energies $E_{\rm F}+\varepsilon$  
($E_{\rm F}$ is the Fermi energy)  are 
coherently scattered into Fermi sea holes of energy 
$E_{\rm F}-\varepsilon$
(and vice versa) at the N-S interface  
if  $\varepsilon$ is  smaller than the superconducting gap $\Delta $.
For   $\varepsilon=0$ 
the Andreev reflected electrons (holes) 
 perfectly retrace their classical trajectories as holes (electrons).
For $\varepsilon>0$ the 
velocities of the quasiparticles are not exactly reversed because of
the   $2\varepsilon$ difference in their energies, but this 
is not considered in most of the literature.

The validity of this widely used retracing approximation is questionable 
for ABs with chaotic normal dots, if the time  
between two Andreev-reflections is significantly  longer than the ergodic time, because
the small misalignment between the velocities of the electrons and holes at 
the N-S interface leads eventually to divergent trajectories. 
One notable exception, in which  the consequences of the non-exact velocity reversal 
were investigated is the paper by Silvestrov et al.~\cite{ref:silvestrov}. 
They studied an  Andreev-billiard with a chaotic normal dot and 
found that even though the motion 
in the normal dot 
could be characterized by non-zero Lyapunov-exponents,
the existence of an adiabatic invariant in the N-S system to  good approximation
confines the electron-hole orbits to tori.
Nevertheless, 
the exponential divergence of nearby trajectories does manifest itself
through a gap in the low energy density of states (DOS) and 
through the eigenfunctions of the system, which were predicted
to exhibit a peculiar localization property,
different from those known  in normal billiards. 
However, no quantum calculations were performed to support the latter finding.  

Recently, Wiersig~\cite{ref:wiersig} has shown that the diffraction   
occurring at the points separating 
the normal and superconducting segments of the dot boundary 
(coined ``critical points'') also plays an important role in the
classical dynamics of ABs. 
Electrons (holes) hitting the boundary at nearby points 
belonging  to different types of  segment  will undergo 
either Andreev or normal reflection and 
will be scattered into 
different quasi-particle  states (ie into holes or electrons).
Owing to the special geometry this diffraction effect was not
observed in Ref.~\cite{ref:silvestrov}.

\begin{figure}[hbt]
\vspace*{-2mm}
\includegraphics[scale=0.38]{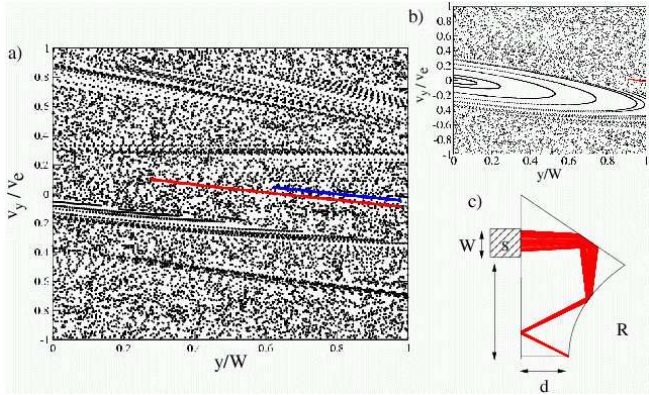}
\vspace*{-5mm}
\caption{(color online)  The Poincar{\'e}  section of the SA billiard 
for $h=0.58$ (a) and $h=0$ (b).
(For other  parameters see \cite{ref:geom-param}).
Each dot represents a starting point 
of an electron trajectory;
the red and blue dots correspond to very elongated regular islands. 
The torus shown with red dots in (a)
is projected onto real space (c).
\label{fig:cha-poinc}}
\end{figure}

In this Letter we show, that in a more general case,
the interplay of the {\em non-exact velocity reversal} and the 
{\em diffraction} at the critical points 
leads to the breakup of the (adiabatic) tori in at least 
certain parts of the phase space 
[see Fig.~\ref{fig:cha-poinc}(a),(b)]
rendering the classical dynamics mixed.
While it has been known for a long time,
that an applied magnetic field can render the ABs 
(weakly) chaotic\cite{ref:andreev-bil,ref:richter},
to our knowledge
the possibility that the dynamics of ABs can be irregular even for 
zero magnetic field has not yet been addressed in the literature.
We argue moreover that the combination of 
these two effects results in a wealth of phenomena 
not only in the classical but also in the quantum dynamics of  ABs.
What makes these N-S hybrid systems additionally 
interesting and novel in the realm of  quantum chaos studies 
is that the wavefunction is spinor and not scalar, 
as in normal billiards.
We calculate, for the first time, the exact quantum eigenfunctions
of a two dimensional AB with chaotic normal dot. Using the Wigner transform of the 
eigenstates we find that most of them can be associated either with the 
regular or the chaotic regions of the energy surface.  

The particular  system  used in our numerical calculations is the 
Sinai-Andreev (SA) billiard\cite{ref:effective_rmt}. It consists  of a Sinai-billiard
shaped normal dot and an attached (infinite) superconducting 
lead (see Fig.~\ref{fig:cha-poinc}(c) for the geometry). 
First, we consider the classical dynamics. 
Fig.~\ref{fig:cha-poinc}(a,b) shows the Poincar{\'e} section (PS) which is  defined as follows. 
The coordinates of each point in PS are the position $y$
 of the electron starting from the N-S interface and the tangential 
velocity component $v_y$ in units of $v_e =v_F
 \sqrt{1+\varepsilon/E_{\rm F}}$, where $v_F$ is the Fermi velocity
($y$ is measured from the lower corner of the N-S interface).  
As  can be observed, the PS resembles the Poincar{\'e} maps of 
generic normal systems  with mixed classical dynamics.
In large regions of the energy surface the motion is {\em chaotic}, while
{\em islands of stability} are preserved only around
such unstable periodic orbits of the isolated normal system,
which hit the N-S interface at right angle. 
We emphasize that in contrast to
Ref.~\cite{ref:silvestrov}, these phase space structures are  
regular islands not only in the adiabatic approximation. 
Depending on the  instability of the orbit  and  the value of 
$\varepsilon/E_{\rm F}$,
the presence of the superconductor can indeed stabilize the motion, 
as  can be  checked by computing the stability matrix of 
the corresponding electron-hole orbit.
Most of these islands are extremely elongated
(see eg  Fig.~\ref{fig:cha-poinc}(a,b), where they are denoted by red and blue dots). 
Moreover, considering the bunch of electron trajectories lying on a particular torus, 
its projection onto the real space 
is ``squeezed down'' away from the N-S interface   [Fig.~\ref{fig:cha-poinc}(c)].
Both  phenomena  are consequences of the exponential divergence of nearby trajectories in the 
 normal dot\cite{ref:silvestrov}. 
For islands centered on the 
least unstable orbits (eg the one at $h=0$) 
these effects are less pronounced [see the large island in Fig.~\ref{fig:cha-poinc}(b)] .

\begin{figure}[hbt]
\vspace*{-3mm}
\includegraphics[scale=0.38]{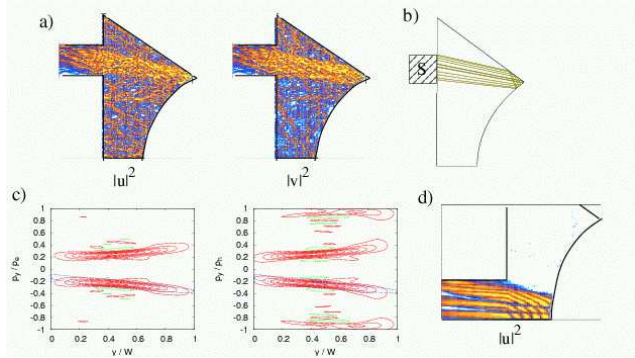}
\vspace*{-3mm}
\caption{(color online)
The probability density of the electron ($|u|^2$) and hole ($|v|^2$) 
components of an intermittent eigenstate 
($\varepsilon/ \Delta_0=0.44652$)
of the SA billiard (a).
One of the  slowly drifting electron-hole orbits during the 
intermittent-like motion (b) 
which form  the  regular strip  around
$v_y/v_e\approx -0.2$ in 
Fig.~\ref{fig:cha-poinc}(a). The orbit   
comprises  electron (red) and hole (green) trajectories. 
The smoothed projections $\mathcal{W}_{\rm P}$
for the electron and the hole components of the eigenstate (c). 
Eight equally spaced 
positive contours (red lines) and one negative contour (green line) are plotted.
The blue curve marks the location of that region of the PS, in which the drifting
electron-hole orbits, like that one in Fig.~\ref{fig:intermit-wf}(b), stay longer than
$\tau_{\rm \scriptscriptstyle H}$ (see text). 
The electron component of a wavefunction
localized onto a quantized torus when $h=0$ (d). The hole component 
is almost identical and thus not shown here. 
\label{fig:intermit-wf}}
\end{figure}

Furthermore, strips of regular, {\em intermittent-like} motions 
can also be observed in 
Fig.~\ref{fig:cha-poinc}(a),
in particular around  $v_y/v_{\rm e}\approx 0.7$, $0.3$, $-0.2$, and $-0.5$.
These regions correspond to such initial conditions for which the electron and 
hole trajectories will  only be slightly  different, but this  nearly periodic orbit 
will slowly drift in  phase space\cite{ref:stone,ref:silvestrov} 
[see also Fig.~\ref{fig:intermit-wf}(b)]. 
This  dynamics  resembles  the intermittent  behavior 
in  normal billiards\cite{ref:dahlqvist}
where this near-integrable motion usually evolves in the vicinity of tori or isolated periodic orbits.
In the present case [Fig.~\ref{fig:cha-poinc}(a)], however,  it is a  consequence of the 
retroreflection mechanism and is therefore a peculiarity of ABs.
Nevertheless, owing to  this drift
either the electron or the hole trajectory will eventually 
reach the edge of the N-S interface and 
then by hitting the normal wall instead of the superconductor,  
escapes to the chaotic sea.

In what follows, we show how the properties  of the classical phase
space leave their fingerprints on the eigenstates
of Andreev billiards. 
First, we  briefly summarize the  quantum treatment of the system.
The spinor wavefunction  $\psi(\mathbf{r})=[u(\mathbf{r}),v(\mathbf{r})]^{T}$ 
(with $u(\mathbf{r})$ electron and $v(\mathbf{r})$  hole components)
of the N-S hybrid systems satisfies the 
Bogoliubov--de Gennes equations 
$\hat{\mathcal{H}}\psi(\mathbf{r})=\varepsilon \psi(\mathbf{r})$,
where $\hat{\mathcal{H}}=\hat{\mathcal{H}}_0 \sigma_z + 
\Delta(\mathbf{r})\sigma_x$, 
and $\hat{\mathcal{H}}_0=-\hbar^2/2m \nabla^2-\mu$
is the single-particle Hamiltonian 
with Dirichlet boundary conditions at the normal walls. Here
$\mu $ is the chemical potential and $\sigma _z$, $\sigma_x$ are Pauli matrices.
We assume  that
the  superconducting pair potential is  
 $\Delta(\mathbf{r})=\Delta_0$  constant inside the lead 
and zero in the N region\cite{ref:beenakker-rev}. 
The calculation of the wavefunction requires two steps: first we obtain the  energy 
levels of the SA billiard  using a quantum mechanically exact secular equation which 
can be derived invoking  the scattering approach of Ref.~\cite{ref:beenakker-rev}. 
The method 
also furnishes us with the wavefunction in the superconducting lead and at 
the N-S interface.
Then the boundary integral method  is employed to find the wavefunction   
in the normal dot\cite{ref:calc-details}. 
We work in the regime $\delta_N\ll E_{\rm T}\ll \Delta_0\ll E_{\rm F}$\cite{ref:beenakker-rev} 
where $\delta_N$ is the mean level spacing of the isolated normal dot and $E_{\rm T}$ is 
the Thouless energy\cite{ref:geom-param}.

The classical-quantal correspondence of phase space structures 
and eigenstates 
 of a given system can be studied with the help of the 
Wigner function\cite{ref:berry-leshouche}. 
In order to compare the Wigner function with the PS,
we  calculated the  projection 
of the Wigner function\cite{ref:robnik} 
 of both the electron and the hole components of the wavefunction onto the PS:
\begin{equation}
\!\mathcal{W}_{\rm P}(y, p_y)
\!=\!\frac{1}{2\pi\hbar}\!\int\!\! {\rm d}Y\,e^{-i p_y Y} \zeta^*(x, y-\frac{Y}{2})\,\zeta(x,y+\frac{Y}{2})\!
\end{equation}
evaluated at the N-S interface (ie $x=0$), $\zeta$ is either 
$u(\mathbf{r})$ or $v(\mathbf{r})$  
and $p_y$ is the parallel (to the interface) component of the momentum. 
The Wigner function
is not positive definite and usually exhibits rapid oscillations that can obscure 
the physical  content. For this reason,  like  in Ref.~\cite{ref:robnik}, we smoothed 
the projection $\mathcal{W}_{\rm P}$ with a Gaussian, which was chosen  narrower 
than the minimum uncertainty Gaussian. 
Note that $\mathcal{W}_{\rm P}$ is symmetric in $p_y$,
since the system is time reversal invariant. 
This symmetry is absent in the Poincar{\'e} map,
because 
incidental electrons (holes) are not taken into account.

In common with  normal billiards, most of the eigenstates of the SA billiards 
can  be classified 
as {\em `chaotic'} or {\em `regular'}~\cite{ref:percival,ref:bohigas,ref:robnik}.
The chaotic eigenstates can further be  subdivided into two
groups: i) ergodic-like, and  ii) scarred states.
Regarding the regular states, they either can be 
associated with the intermittent-like regions 
of the phase space or with quantized tori. 
We now consider each case.

\begin{figure}[hbt]
\vspace*{-3mm}
\includegraphics[scale=0.4]{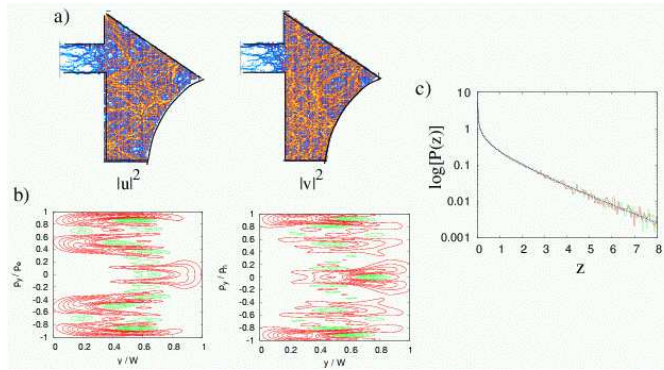}
\vspace*{-5mm}
\caption{(color online) The probability densities of the electron ($|u|^2$) and hole ($|v|^2$) 
components of a chaotic eigenstate ($\varepsilon/ \Delta_0=0.060395$) 
of the SA billiard (a). 
The corresponding 
smoothed projections of the Wigner function onto the PS (b)
in units of $y/W$ and $p_y/p_e$, $p_y/p_h$ for the electron and hole components respectively.
Here $p_e=m v_e$ and $p_h(\varepsilon)=p_e(-\varepsilon)$. Eight equally spaced 
positive contours (red lines) and five negative contours (green lines) are plotted. 
The distribution of the scaled probabilities  $z_e$ (red line), $z_h$ (green line) and the
 Porter-Thomas distribution  (blue line) (c). 
\label{fig:chaotic-wf}}
\end{figure}

For ergodic-like eigenstates (an example  is shown in Fig.~\ref{fig:chaotic-wf}(a)) 
both the electron and the hole
components of the wavefunction seems to cover the normal dot 
in a roughly uniform way.
Nevertheless, they display  different interference pattern which 
translates also into
 the corresponding projections of the Wigner functions. 
Examining  Fig.~\ref{fig:cha-poinc}(a) and Fig.~\ref{fig:chaotic-wf}(b) 
one can see that  $\mathcal{W}_{\rm P}$ has high amplitude
in such regions, which corresponds to chaotic regions in the PS. 
This observation is reinforced by the good agreement between the probability 
distribution of the scaled local densities  $z_e(\mathbf{r})$, $z_h(\mathbf{r})$ 
and the  Porter-Thomas distribution 
$\mathcal{P}(z)=1/ \sqrt{2\pi z}\exp(-z/2)$ shown in Fig.~\ref{fig:chaotic-wf}(c).
Here $z_{e}(\mathbf{r})=(\mathcal{A}/\mathcal{N}_e) |u(\mathbf{r})|^2$, 
 where $\mathcal{N}_e=\int |u(\mathbf{r})|^2 {\rm d}^2 \mathbf{r}$, and the integration 
is performed  over the area $\mathcal{A}$ 
of the normal dot (an analogous definition applies for the hole 
density $z_h(\mathbf{r})$). 
Both the electron and the hole components of  chaotic eigenstates of ABs
can thus be considered, similarly to the eigenstates of chaotic 
normal billiards\cite{ref:berry-leshouche, ref:ishio}, as being a 
 superposition  of infinite number of plane waves with fixed wave number, 
but with random directions and amplitudes. 

The old wisdom of the quantum chaos: 'the scars are scarce' seems 
to hold also for ABs.
\begin{figure}[hbt]
\vspace*{-3mm}
\includegraphics[scale=0.38]{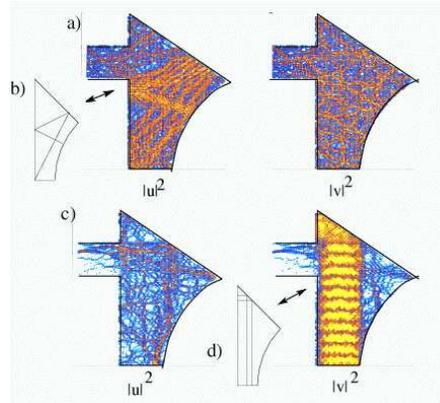}
\vspace*{-3mm}
\caption{ (color online) The probability densities of the electron 
($|u|^2$) and hole ($|v|^2$) components of a scarred eigenstate 
($\varepsilon/ \Delta_0=0.88323$) of the SA billiard (a) and the corresponding 
 unstable periodic orbit (b).
The probability densities of a decoupled 
state at $\varepsilon/ \Delta_0=0.42901$ (c) and the marginally stable
periodic  orbit family over which the hole component shows  enhancement (d).
 \label{fig:scarred-wf}}
 \end{figure}
\vspace*{-1mm}
Out of  99 eigenstates 
of the system with geometrical parameters listed in the
caption of Fig.~\ref{fig:cha-poinc}(a), only 2 can be 
classified as scarred, one of them  is shown in Fig.~\ref{fig:scarred-wf}(a) along with 
the periodic orbit that scars the electron component (Fig.~\ref{fig:scarred-wf}(b)). 
We call these states 'genuine Andreev-scarred'  to distinguish them from 
the other type of states which are decoupled 
from the superconductor\cite{ref:florian} and in certain cases 
also show enhancement over periodic orbits 
(see  Figs.~\ref{fig:scarred-wf}(c) and (d)). 
In the case of decoupled eigenstates the amplitude of the wavefunction 
at the N-S interface is small and the probability 
of finding the quasiparticle in either the electron or hole state
inside the dot is enhanced. 
For the example  shown in Fig.~\ref{fig:scarred-wf}(d), 
 integrating $|v(\mathbf{r})|^2$
over the area of the billiard one finds that this probability is 91\% whereas 
only 7\% for the electron component 
(and 2\% corresponds to the probability of quasiparticles in the superconductor).
In contrast, for the eigenstate shown in  Fig.~\ref{fig:scarred-wf}(a) 
the wavefunction has an apparently finite value at the N-S interface and the 
probabilities are 56\% (electrons) and 31\% (holes).

Regarding the regions of regular motion, we found that both the 
intermittent-like parts of the phase space and the regular islands 
can give rise to quantum eigenstates. 
The electron and the hole components  of regular eigenstates   
display very  similar interference patterns  and the probabilities 
of the electron and hole state in the dot have close values.
Furthermore, 
the localization of both the eigenfunction and the corresponding  
smoothed  $\mathcal{W}_{\rm P}$  onto the underlying  phase space
structure can  be observed.
As for the  islands of stability shown in Fig.~\ref{fig:cha-poinc}(a), 
they enclose a  very small area and thus  for the numerically 
accessible wavelengths they
are not resolved quantum mechanically. 
A much more amenable system for studying 
the correspondence between  quantum eigenstates and a regular island
can be obtained by attaching the 
superconducting lead at $h=0$, 
since in this case the emerging regular island is larger
[see Fig.~\ref{fig:cha-poinc}(b)].
Our findings on the eigenfunction 
properties listed above are based on studying this case [see 
also Fig.~\ref{fig:intermit-wf}(d)], which is, however,
not treated here in more detail  for the lack of space.
Finally, we found that as  expected, 
those intermittent-like regions support quantum eigenstates, which  
correspond to  staying times  longer  than the Heisenberg time 
$\tau_{\rm\scriptscriptstyle H}=2\hbar/ \delta_{\rm N}$. 
An example is shown in Fig.~\ref{fig:intermit-wf}(a),
while Fig.~\ref{fig:intermit-wf}(c) shows that  the 
smoothed  $\mathcal{W}_{\rm P}$  also has high amplitude  in the corresponding region.

In summary, we investigated the quantal-classical correspondence in 
Andreev billiards with a chaotic normal dot. 
We showed that in contrast to the common belief, 
the interplay of critical points and non-exact velocity 
reversal can 
render the classical dynamics non-integrable
 even for zero magnetic field.   
It was also shown that the eigenstates of the 
 system can be classified as chaotic or regular corresponding to 
different regions of  phase space.
This implies that while the retracing approximation has been proved 
to be useful in understanding the energy dependence of the 
density of states\cite{ref:melsen, ref:florian},
it may  not be adequate when addressing the properties of 
individual eigenstates. 
Experimentally, the eigenstates of the Andreev billiards 
might be studied using scanning tunneling probe, and 
those which comprise quasiparticle 
density enhancement in certain part of the dot should be discernible
from the ergodic ones.

We would like to thank C.~W.~J.~Beenakker, P.~G.~Silvestrov and H.~Schomerus
for useful discussions.
This work is supported 
by E. C. 
Contract No. MRTN-CT-2003-504574, EPSRC,  
the Hungarian-British TeT, 
and the Hungarian  Science Foundation OTKA TO34832.


\begin{thebibliography}{99}
\bibitem{ref:calc-details}
Details of the calculations will be given elsewhere.
\bibitem{ref:geom-param}
The value of $\varepsilon/ E_{\rm F}$ is $0.0069$ for Figs.~\ref{fig:cha-poinc}(a,b).
 Throughout this paper the number of open channels in the lead is
$k_{\rm F} W/ \pi=25.51$ ($k_{\rm F}$ is the Fermi wavenumber), while 
 $\Delta_0/E_{\rm F}=0.015$ and 
$W=0.21$, $R=0.8$.
\end{thebibliography}
\end{document}